# Role of direct laser acceleration in energy gained by electrons in a laser wakefield accelerator with ionization injection

J L Shaw[1], F S Tsung[2], N Vafaei-Najafabadi[1], K A Marsh[1], N Lemos[1], W B Mori[1,2] and C Joshi[1]

[1]Department of Electrical Engineering & [2]Department of Physics and Astronomy, University of California Los Angeles, 405 Hilgard Ave, Los Angeles, California, United States 90095

Email: jshaw05@ucla.edu

**Abstract.** We have investigated the role that the transverse electric field of the laser plays in the acceleration of electrons in a laser wakefield accelerator (LWFA) operating in the quasi-blowout regime through particle-in-cell code simulations. In order to ensure that longitudinal compression and/or transverse focusing of the laser pulse is not needed before the wake can self-trap the plasma electrons, we have employed the ionization injection technique. Furthermore, the plasma density is varied such that at the lowest densities, the laser pulse occupies only a fraction of the first wavelength of the wake oscillation (the accelerating bucket), whereas at the highest density, the same duration laser pulse fills the entire first bucket. Although the trapped electrons execute betatron oscillations due to the ion column in all cases, at the lowest plasma density they do not interact with the laser field and the energy gain is all due to the longitudinal wakefield. However, as the density is increased, there can be a significant contribution to the maximum energy due to direct laser acceleration (DLA) of those electrons that undergo betatron motion in the plane of the polarization of the laser pulse. Eventually, DLA can be the dominant energy gain mechanism over acceleration due to the longitudinal field at the highest densities.

## 1. Introduction

As laser wakefield acceleration (LWFA) becomes a more mature field of research, there is an increasing emphasis on perfecting the different ways in which electrons can be injected into the wake [1-5]. To gain a better understanding of these electron injection processes and the subsequent acceleration within the wake, one frequently resorts to particle-in-cell (PIC) code simulations. For instance, these simulations have been utilized to verify the phenomenological scaling laws for the maximum electron energy gain of the electrons in the so-called ideal blowout regime in the case of self-injection of electrons into the wake [6]. The LWFA is said to be in the blowout or bubble regime when the laser pulse is intense (normalized vector potential $a_0 = \frac{eA}{mc^2} > 4$), short ($c\tau < \frac{2\sqrt{a_0}c}{\omega_p}$), and matched to the plasma density (laser spot size $w_0 \sim$ the blowout radius $R_b$) to completely expel all the plasma electrons. Here, c is the speed of light, $\tau$ is the pulse length, $\omega_p = \sqrt{\frac{e^2 n}{m\epsilon_0}}$ is the plasma frequency, e is the charge of the electron, n is the plasma density, m is the mass of the electron, and $\epsilon_0$ is the permittivity of free space. However, many, if not most, experiments to date have employed laser beam parameters where considerable longitudinal and transverse laser pulse evolution is necessary for the injection of electrons into the wakefield [7]. This range of experimental parameters has made the comparison of experimental data with the scaling laws difficult. To circumvent the need for this laser pulse evolution, the ionization injection technique [2-3, 8-9] can be employed for injecting the electrons into the wake. For the range of plasma densities (mid-$10^{18}$ to a few $10^{19}$ cm$^{-3}$) and laser pulse durations (35-45 fs full-width, half-maximum (FWHM)) that are typically used in many experiments, the laser pulse may occupy only a fraction of the first wavelength of the wake oscillation (the accelerating bucket) at low plasma densities, whereas at the highest plasma density, the same duration laser pulse may fill the entire first bucket. It is therefore important to understand the role of not only the longitudinal electric field of the wake, but also the other fields—namely the transverse fields of the ion column and of the laser itself—play in determining the ultimate energy gained by the electrons. The process by which electrons that undergo betatron motion in the plane of polarization of the laser pulse gain energy directly from the laser field itself is called direct laser acceleration (DLA) [10-11]. The purpose of this paper is to demonstrate through two-dimensional (2D) PIC code simulations that in this regime of laser and plasma parameters, one can transition from pure LWFA at low densities to DLA at high plasma densities.

## 2. Background

To set the stage, we first review how electrons gain energy from plasma accelerators [12] and from the laser field itself (DLA). Since plasma accelerators can operate in different regimes depending on laser and plasma parameters, we begin with intense, but long, laser pulses propagating in the plasma. If the laser pulse is much longer than the plasma wavelength, then the plasma accelerator is said to operate in the self-modulated LWFA (SMLWFA) [13] and/or the Raman forward scattering (RFS) regime [14-15]. Since the laser pulse is many plasma periods long, the SMLWFA or FRS process requires laser pulses with $a_0 \sim 1$. If the $a_0$ is much higher than this, then electron cavitation [16] can compete with plasma wave generation and produce an ion column, which is a region devoid of plasma electrons. Some electrons can then be pre-accelerated either by the longitudinal field of a plasma wave excited by the laser before all the plasma electrons are blown out or by the fields of a surface wave surrounding the ion cavity. These electrons can now oscillate under the combined influence of the laser field and the ion field and gain more energy in a process that was termed "direct laser acceleration" [17-18].

If the length of the laser pulse is on the order of the plasma wavelength and if the laser intensity is insufficient to cause a complete blowout of the plasma electrons, then the wake generation is said to be in the forced wakefield regime [19-21]. In this regime, the pulse creates an initial plasma wave, which then shortens the frequency-modulated laser pulse through group velocity dispersion. The laser pulse forms an "optical shock" with a steep front edge, which can then drive an even stronger wake. Some LWFA experiments to date have been carried out in this forced wakefield regime, and in the absence of theoretical models, the results are either compared with a LWFA operating in the blowout regime or with PIC code simulations [22].

When the laser pulse is shorter than the plasma wavelength but intense enough to fully expel all electrons from the laser axis (electron blowout), the LWFA can operate in the ideal blowout regime where the resulting wake has a "bubble" shape. This regime is characterized by a laser driver with an $a_0 \geq 4$, a pulse length less than $\frac{2\sqrt{a_0}}{\omega_p}$, and a normalized $w_0$ that satisfies $k_p R_b \cong k_p w_0 = 2\sqrt{a_0}$ where $k_p = \omega_p/c$ is the plasma wavenumber. However, simulations show that even if $2 \leq a_0 \leq 4$, the LWFA can still operate close to the blowout regime (i.e. the quasi-blowout regime) if the other conditions hold.

The fully blown out wakefield structure has a desirable transverse and longitudinal field structure for generating a high-quality electron bunch, but it also has all the conditions needed for DLA if there is an overlap between the trapped electrons and the transverse electric field of the laser pulse. Provided that there is another electric or magnetic field that gives the electrons a velocity component in the direction of the laser polarization, DLA can accelerate electrons in the longitudinal direction using the transverse field of a laser [10-11]. In a LWFA operating in the blowout regime, the ion column acts as a very strong wiggler. Trapped electrons that are being accelerated by the wake undergo betatron oscillations in response to the transverse electric field of the ion column [23-24]. The length of the bucket scales inversely with the square root of the plasma density [6]. Therefore, for a given laser pulse length, if the bucket is shortened by increasing the plasma density, a LWFA can be configured such that some of the trapped electrons undergo betatron oscillations in the plane of polarization of the laser's electric field soon after they are trapped and directly exchange energy with the laser field (DLA).

DLA can not only increase the energy of the accelerating electrons in a LWFA, but it can also lead to an increase in the amplitude of the betatron oscillations. This increased amplitude can lead to very high energy photons being radiated by the oscillating electrons [25]. In fact, even though the role of DLA in the energy gained by electrons in a LWFA has not been demonstrated directly, it has been inferred indirectly from the MV photon emission observed in the forward direction in a recent LWFA experiment [26].

In the single-particle description of an electron oscillating in a plane electromagnetic wave in the presence of a static electric field, the resonance condition for an increase in the transverse momentum of the electrons due to the laser field is given by [10-11]

$$N\omega_\beta = \left(1 - v_\parallel / v_\phi\right) \omega_0 \quad (1)$$

where N assumes integer values, $\omega_\beta = \frac{\omega_p}{\sqrt{2\gamma}}$ is the betatron frequency, $\gamma$ is the Lorentz factor of the electrons, $v_\parallel$ is the velocity of the electron in the longitudinal direction, $v_\phi$ is the phase velocity of the electromagnetic wave (i.e. laser), and $\omega_0$ is the frequency of the electromagnetic wave (i.e. laser). This transverse momentum can then be converted to longitudinal momentum via the $\vec{v} \times \vec{B}$ force [18, 25]. The process of DLA in a LWFA is therefore analogous to the acceleration mechanism in an inverse free electron laser (IFEL) [27-30] except that in DLA, the ion cavity of the LWFA replaces the magnetic wiggler of an IFEL. However, unlike in an IFEL, sustained resonance for DLA is more difficult to design because in the latter case, the normalized undulator strength $K \gg 1$ and the energy of the electrons, the betatron wavelength, and the amplitude of the betatron oscillation are continuously and rapidly changing.

We have therefore used 2D PIC code simulations to investigate the relative importance of LWFA and DLA in the typical laser and plasma parameters used in many experiments. For a constant plasma density, we vary the pulse length to demonstrate that the laser pulse must overlap the trapped electrons for DLA to be present. For a constant laser pulse duration, we vary the plasma density such that at lowest densities the laser pulse occupies only a fraction of the accelerating bucket, whereas at the highest density the same duration laser pulse fills the entire first bucket. If DLA is present as an additional acceleration mechanism, the accelerated electron beams should have higher maximum energies than is expected from the wake acceleration alone. Furthermore, the DLA contribution to the total energy gain should increase with the plasma density because at higher densities, the decreased bucket length will cause the trapped electrons to overlap with the more intense regions of the laser, leading to higher DLA contributions.

## 3. Simulations

The simulations were carried out using the 2D PIC code OSIRIS [31] in the speed-of-light frame with the ADK [32] ionization package. The simulation box size ranged between 36 x 54 $c/\omega_p$ = 67.8 x

101.6 µm where $c/\omega_p$ was 1.88 µm for the lowest-density (8 x $10^{18}$ cm$^{-3}$) simulation and 80 x 94 $c/\omega_p$ = 87.0 x 102.2 µm where $c/\omega_p$ was 1.09 µm for the highest-density (2.4 x $10^{19}$ cm$^{-3}$) simulation. The number of grid points was varied from 2494 x 452 = 1.13 x $10^6$ to 3200 x 456 = 1.46 x $10^6$ for the 8 x $10^{18}$ to 2.4 x $10^{19}$ cm$^{-3}$ simulations, respectively. Sixteen particles per cell were used for a total of ~18-23 million particles for the 8 x $10^{18}$ to 2.4 x $10^{19}$ cm$^{-3}$ simulations, respectively. The transverse resolution $k_p dx_\perp$ was between 0.12 and 0.21 for the 8 x $10^{18}$ to 2.4 x $10^{19}$ cm$^{-3}$ simulations, respectively. The longitudinal resolution in the direction of laser propagation was $k_0 dx_\parallel = 0.21$ in all cases. At the outset, it should be stated that the transverse evolution of the laser pulse, the trapping efficiency of the ionization-injected electrons, and the degree of electron blowout will be somewhat different in three-dimensional simulations compared to the 2D simulations presented here. Therefore, the results of the cases considered in this paper are meant to qualitatively highlight the relative importance of the physical effects.

In the simulations, the initial neutral gas profile was trapezoidal with a uniform gas region that was 1800 µm long and with 100 µm linear entrance and exit ramps on either side. The neutral gas was comprised of 99.9% helium and 0.1% nitrogen. The ionization of the K-shell electrons of the nitrogen served to inject charge into the wake formed predominantly by the ionization of helium atoms as soon as it is formed. The plasma density was varied from 8 x $10^{18}$ cm$^{-3}$ to 2.4 x $10^{19}$ cm$^{-3}$. The central laser wavelength was 815 nm, and the laser pulse was focused in the center of the density up-ramp at the entrance of the plasma (see figure 1).

For the first two cases described below, the plasma density was 8 x $10^{18}$ cm$^{-3}$, and the FWHM laser pulse durations $\tau_p$ were 30 fs and 45 fs. The laser power P was 6.4 TW or $P/P_{crit}$ was 1.8, where $P_{crit}$ [GW] $\cong 17.4 \frac{\omega_0^2}{\omega_p^2}$ is the critical power for self-focusing. The laser spot size was 6.7 µm (half-width, half-max of intensity) giving a vacuum $a_0$ of 2.1. The laser's temporal profile was a polynomial fit to a Gaussian. In the $\tau_p = 30$ fs simulation, which acted as the control case, the laser pulse did not significantly overlap the trapped electrons, and therefore one does not expect a DLA contribution to the wakefield acceleration. The $\tau_p = 45$ fs simulation models the case where the back (falling intensity) of the laser pulse overlaps the trapped electrons, and therefore one might expect some contribution from DLA to the maximum energy gain.

First, we ran the simulations to obtain the total energy gained by the K-shell nitrogen electrons and identified the 40 electrons that had gained the most energy. We subsequently ran the same simulation again and tracked the trajectories of these 40 electrons. Tracking these electrons records their velocity, momentum, and position for each time step. The energy gained by the electrons due to the longitudinal field $\vec{E}_\parallel$ was calculated for each tracked particle by evaluating the integral

$$e \int_0^t \vec{v}_\parallel \cdot \vec{E}_\parallel dt' \qquad (2)$$

for each time step of the simulation. The maximum possible DLA contribution was calculated by evaluating the integral

$$e \int_0^t \vec{v}_\perp \cdot \vec{E}_\perp dt' \qquad (3)$$

at each time step of the simulation. In this equation, $\vec{v}_\perp$ is the velocity of the tracked electron in the plane of the laser polarization, and $\vec{E}_\perp$ is the transverse field of the laser. Note that no direct energy exchange between an electromagnetic wave and a charged particle is possible unless the electron has a velocity component in the direction of the plane of the polarization of the laser. As stated earlier, this transverse velocity component comes from the betatron motion of the electrons in the ion cavity of the wake. In DLA, this perpendicular momentum component is then amplified by the transverse field of the laser.

We first describe the results of the 30-fs-laser-pulse-width "control" simulation in figure 1(a). For these laser and plasma parameters, complete expulsion of the plasma electrons, which characterizes the ideal blowout regime, was not observed. Only 73% of the electrons were evacuated resulting in a residual, on-axis plasma electron density of 2.1 x $10^{18}$ cm$^{-3}$ within the first bucket. The energy gain due to the wakefield is approximately linear with distance, and the maximum energy gain is limited by dephasing [6] and not pump depletion [33]. The maximum energy gain (black curve) of these electrons is 234 MeV, which occurs at 1440 µm into the plateau region of the plasma. This location is 660 µm beyond the predicted dephasing length $L_{dephase} \cong \frac{c}{c-v_\phi} R_b \cong \frac{2}{3}\frac{\omega_0^2}{\omega_p^2} R_b$ = 780 µm [6], which is primarily due to the lengthening of the first bucket by the non-optimal 30-fs-long laser pulse. Consequently, the maximum electron energy exceeds the dephasing-length-limited energy gain of 150 MeV predicted by $\Delta E \cong \frac{2}{3}mc^2 \frac{\omega_0^2}{\omega_p^2} a_0$ [6].

Recall that in this case, there is no significant overlap between the trapped electrons and the laser field at the back of the bucket where trapping occurs; thus, no DLA is expected. Figure 1(a) indeed shows this to be the case. The wakefield contribution (red curve) is the same as the total energy gain (black curve). The DLA contribution (blue curve) in this case is negligible. Therefore, for the LWFA regime where the laser pulse does not overlap the trapped electrons, the wakefield is the only source of the energy gain for these highest-energy electrons as expected.

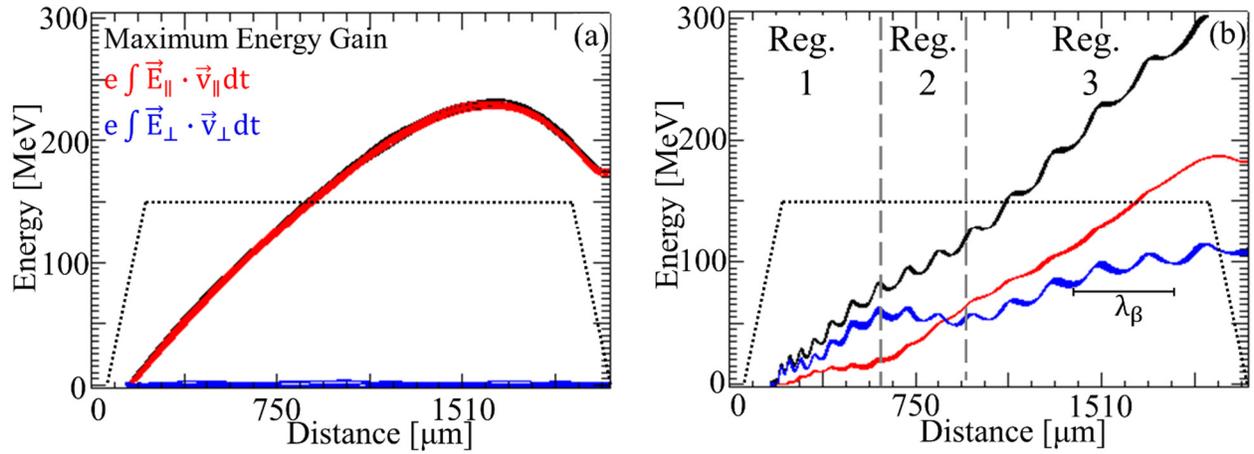

Figure 1: (a) Plot of the maximum energy gain (black curve), LWFA contribution (red curve calculated using (2)), and DLA contribution (blue curve calculated using (3)) for the tracked electrons in the 2D OSIRIS simulation of a 30 fs laser pulse with an $a_0$ of 2.1 propagating through 1800 µm of 8 x $10^{18}$ cm$^{-3}$ plasma. The dotted black curve marks the plasma density profile. (b) Plot of the maximum energy gain, longitudinal field contribution, and DLA contribution as a function of the distance for a 45 fs laser pulse with an $a_0$ of 2.1 propagating in the same plasma as (a). In regions 1 and 3, the electrons are gaining energy from the transverse laser field. In region 2, the electrons are slowly losing energy to the transverse laser field. Note that all the electrons in (b) oscillate nearly in phase with one another.

The results of the energy gain by the highest-energy electrons for the case of the 45 fs laser pulse are shown in figure 1(b). In this simulation, complete blowout of the plasma electrons is again not seen. The typical residual on-axis plasma density was 3.2 x $10^{18}$ cm$^{-3}$, which corresponds to 60% blowout. Figure 1(b) shows that the maximum energy gain (black curve) in the 45 fs case is 302 MeV, which is

29% higher than the energy gain in the control simulation (30 fs) with no DLA. The maximum longitudinal field contribution (red curve) is 187 MeV and occurs 11 μm into the density down-ramp. There is negligible energy gain in this 11 μm of density down-ramp. The acceleration due to $\vec{E_\parallel}$ persisted for a distance more than twice the dephasing length of 780 μm predicted by the scaling for the ideal blowout regime [6]. There are two primary factors that contribute to this increased dephasing length. First, the 45-fs pulse elongates the bucket even more than the 30-fs pulse. Second, the energy gain due to DLA (blue curve) reduces the longitudinal velocity of the electrons, effectively further increasing their dephasing length compared to the 30 fs simulation where no DLA is observed. For this case, the DLA contribution (calculated using (3)) is 115 MeV. Comparing the $\vec{E_\parallel}$ contribution of 204 MeV to the maximum energy gain of 302 MeV shows that the longitudinal field acceleration only accounts for approximately 2/3 of the maximum energy gain.

In figure 1(b), one can clearly see a sinusoidal structure on both the DLA and the maximum energy gain. This structure is due to the betatron motion. The betatron wavelength $\lambda_\beta$ was measured for each oscillation in the simulation and plotted against the average electron energy over that period in figure 2. To put some theoretical bounds on this plot, the theoretical $\lambda_\beta$ is plotted for full blowout (blue curve) and for 60% blowout (red curve), which is representative of the blowout seen in the simulation. Figure 2 shows that the measured data falls within the theoretical expectations and verifies that the sinusoidal structure on the maximum electron energy and the DLA contribution curves in figure 1(b) is caused by the betatron motion of the trapped electrons in the ion column. As the energy of the electrons increases, so does the oscillation wavelength as might be expected since $\lambda_\beta = \frac{2\pi c \sqrt{2\gamma}}{\omega_p}$.

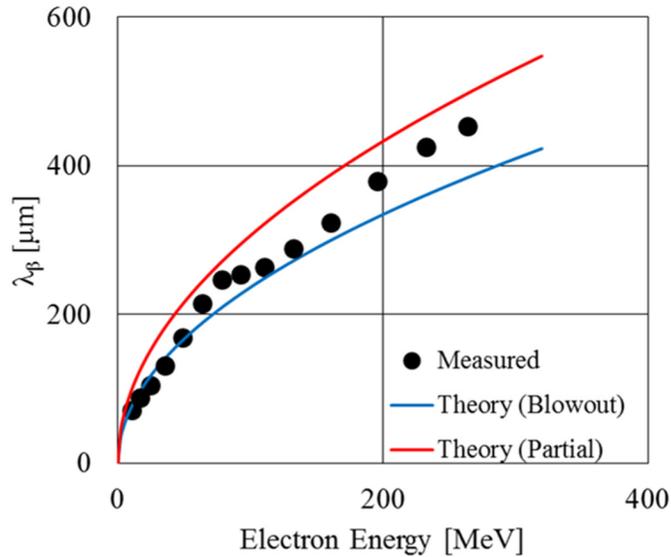

Figure 2: Plot of the betatron wavelength measured in the $\tau_p = 45$ fs simulation (black dots) versus the average electron energy over that oscillation. The blue curve shows the theoretical betatron wavelength calculated using $\lambda_\beta = \frac{2\pi c \sqrt{2\gamma}}{\omega_p}$ assuming complete blowout. The red curve shows the theoretical betatron wavelength accounting for the partial (60%) blowout seen in the simulation.

A closer examination of the DLA contribution in figure 1(b) reveals that the DLA contribution does not steadily increase. Rather, it initially increases up to 620 μm (region 1), then slightly falls until 930 μm (region 2), and then begins rising again (region 3). This behavior of the DLA contribution suggests that the electrons are coming in and out of DLA resonance and/or that these electrons are bunched and return energy to the laser field depending on their relative phase with respect to the laser field. Soon after the electrons are born, they begin to gain energy from the transverse laser field as soon as they begin oscillating in the ion column. In figure 3, we show the transverse momentum of the tracked electrons versus the longitudinal energy gain from DLA. One can clearly see that the transverse momentum is being converted into longitudinal momentum for regions 1 and 3 where the DLA contribution increases in figure 1(b). In region 2, the electrons are repeatedly gaining and losing both transverse and longitudinal momentum.

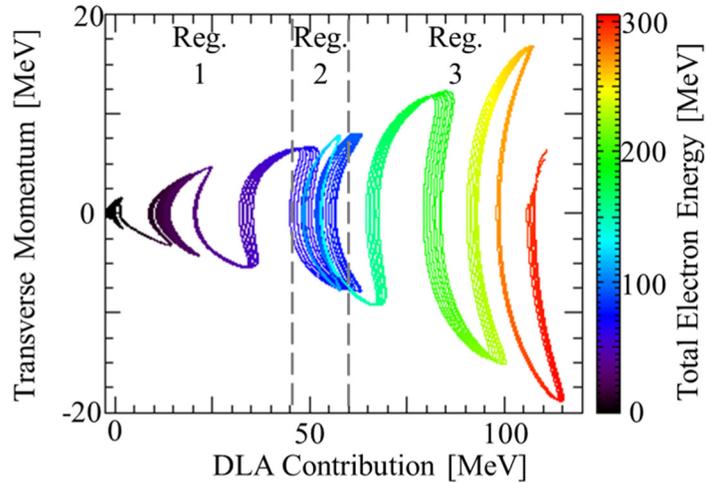

Figure 3: Plot of the transverse momentum versus the DLA contribution in the simulation with $\tau_p$ =45 fs and a plasma density of 8 x $10^{18}$ cm$^{-3}$. The color bar indicates the total electron energy. Regions 1, 2, and 3 correspond to the regions marked in figure 1(b).

One would expect that for the regions where the DLA contribution is increasing, the resonance condition given by (1) is satisfied. We use a sample location at 547 µm, where the transverse momentum of the tracked electrons is zero, in the first region of DLA gain to show that is indeed the case. At this location in the simulation, the longitudinal velocity of the tracked electrons is known, and the betatron frequency, phase velocity, harmonic, and laser frequency sampled by the tracked electrons are all measured. The betatron frequency was measured by finding the betatron wavelength at 547 µm using a best-fit through the betatron wavelengths measured in figure 2 and then using that value to calculate $\omega_\beta$. Figure 4 shows the stacked lineouts of the constant-phase fronts of the laser pulse that were used to determine the phase velocity sampled by the tracked electrons. The phase velocity of $v_\phi/c$ = 1.00164 was determined by measuring the slope of the phase front at the electron location of 547 µm. Since the laser frequency at the back of the laser pulse is blue-shifted due to photon acceleration, a Wigner transform of the laser pulse (figure 5) was taken to determine the laser frequency bandwidth (2.36-2.53 x $10^{15}$ Hz) sampled by the electrons at 547 µm. The resonance condition given by (1) can be rewritten as

$$\frac{v_\parallel}{c} = \frac{v_\phi}{c}\left(1 - N\frac{\omega_\beta}{\omega_0}\right) \qquad (4)$$

Therefore, the longitudinal velocity of the electrons that satisfies the DLA resonance condition for the observed laser bandwidth can be calculated and then used to find the energy of the electrons. For the values measured for the tracked electrons at 547 µm, that energy is 65 MeV for the fundamental harmonic. The energy of the tracked electrons at this location is 65 MeV, which is in agreement with the calculated energy as expected since these electrons are in a region where they are gaining energy directly from the laser field.

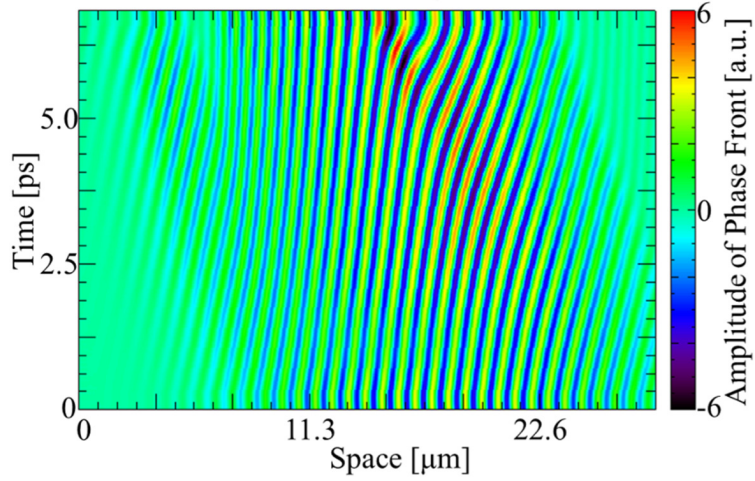

Figure 4: Stacked lineouts of the constant-phase fronts of the laser pulse in the $\tau_p$=45 fs simulation. If the slope of a phase front is perfectly vertical, then $v_\phi = c$; where they are tilted forward, $v_\phi > c$.

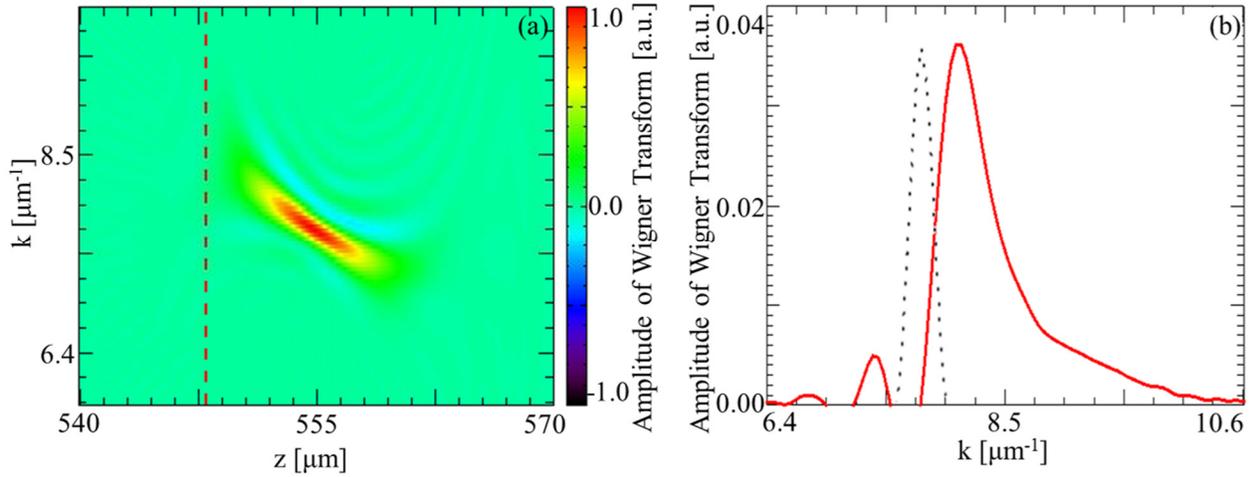

Figure 5: (a) Wigner transform of the laser pulse at the time that the tracked electrons were at 547 µm. The red dashed line marks the location of the tracked electrons. (b) Lineout of the Wigner transform (solid red line) taken at the location of the tracked electrons (marked by red dashed line in (a)). Initial laser spectrum shown by the dotted black line.

Since the total possible DLA contribution is proportional to the transverse field of the laser as shown in (3), it should increase if the electrons are overlapped with the more intense regions of the laser pulse as would be the case if the plasma density is increased, which causes the bucket length to decrease, for a fixed laser pulse length. To investigate this hypothesis, five simulations were carried out that had parameters identical to the control simulation except that the plasma density was varied between 8 x $10^{18}$ cm$^{-3}$ to 2.4 x $10^{19}$ cm$^{-3}$ so that different regions of the $\tau_p$ = 30 fs laser pulse would overlap the trapped electrons. The maximum possible $e \int \vec{v}_\perp \cdot \vec{E}_\perp dt$ and $e \int \vec{v}_\parallel \cdot \vec{E}_\parallel dt$ contributions were measured in each simulation, and the results are shown in figure 6. The trend seen in figure 6 confirms that the maximum possible DLA contribution (e $\int \vec{v}_\perp \cdot \vec{E}_\perp dt$) increases with plasma density in this regime of LWFA. Figure 6 also shows that the maximum possible e $\int \vec{v}_\parallel \cdot \vec{E}_\parallel dt$ contribution decreases with increasing plasma density, which suggests that the contribution of DLA can exceed that of the wake. Therefore, taken as a whole, figure 6 shows that for a fixed laser pulse length, the dominant energy gain mechanism changes from LWFA to DLA as the plasma density is increased.

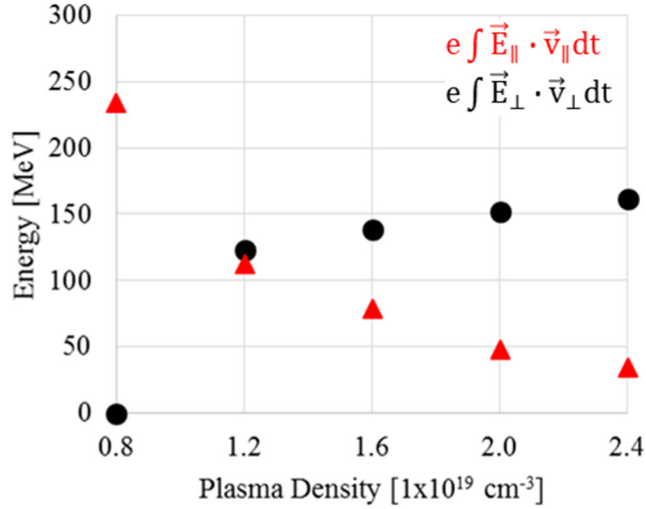

Figure 6: Plot of the maximum possible $\vec{E}_\perp$ and $\vec{E}_\parallel$ contributions versus the plasma density for a fixed laser pulse length of 30 fs. The black dots mark the maximum possible DLA contribution e $\int \vec{E}_\perp \cdot \vec{v}_\perp dt$ in each simulation, which occurs at 1163, 832, 611, and 513 µm into the constant density regions of the $1.2 \times 10^{19}$, $1.6 \times 10^{19}$, $2.0 \times 10^{19}$, and $2.4 \times 10^{19}$ cm$^{-3}$ simulations, respectively. The red triangles mark the maximum longitudinal field contribution, which occurs at 1457, 825, 561, 303, and 269 µm into the constant density regions of the $8 \times 10^{18}$, $1.2 \times 10^{19}$, $1.6 \times 10^{19}$, $2.0 \times 10^{19}$, and $2.4 \times 10^{19}$ cm$^{-3}$ simulations, respectively.

Close examination of the simulation done at $2.4 \times 10^{19}$ cm$^{-3}$ also revealed a unique interplay between the acceleration due to the longitudinal and transverse fields. At this high density, the electron bunch quickly dephases as seen in figure 7, so the longitudinal field contribution to the maximum energy gain is small. Beyond dephasing, the $\vec{E}_\parallel$ contribution and the total maximum energy both begin to decrease as the electrons begin to dephase with respect to the wake. However, at this density, there is a significant DLA contribution that leads to maximum energy gains that exceed the dephasing-length-limited energy [6]. The blue curve in figure 7 shows that the total possible DLA contribution calculated using (3) is 149 MeV. The DLA contribution saturates at 740 µm into the simulation as the laser field begins to pump deplete. Therefore, even though the longitudinal field contribution calculated using (2) begins decreasing from its maximum value of 35 MeV due to dephasing, the maximum energy gain still increases due to the DLA contribution.

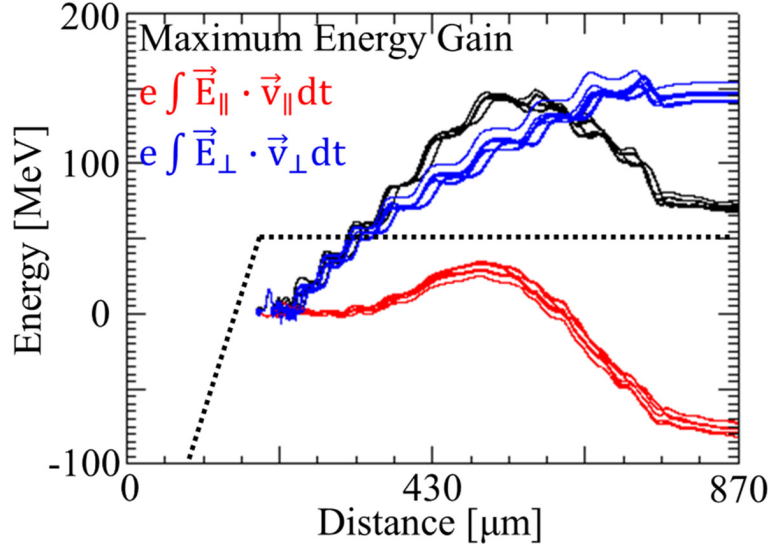

Figure 7: Plot of the maximum energy gain (black curve), $\vec{E}_{\parallel}$ contribution (red curve), and $\vec{E}_{\perp}$ (DLA) contribution (blue curve) for a 2D OSIRIS simulation of a 30 fs laser pulse with an $a_0$ of 2.1 propagating through 680 μm of 2.4 x $10^{19}$ cm$^{-3}$ plasma. The dotted black curve marks the plasma density profile.

In figure 7, the $\vec{E}_{\parallel}$ contribution starts to increase much later than the DLA contribution. This behavior is due to the fact that in ionization injection, the inner K-shell electrons from the nitrogen atoms are born at the highest-intensity regions of the laser pulse [2]. In the relatively high density of this simulation, the highest-intensity region of the laser pulse roughly coincides with the zero of the longitudinal wakefield found at the center of the first bucket. Further, the wakefield is heavily modulated by the laser at this plasma density. Therefore, these electrons are born in a region of the wake where the longitudinal field averages to ~ 0 and do not begin to gain energy from the wake as they drift backward in the bucket until they experience a net accelerating longitudinal field.

At 620 μm, which is 430 μm into the constant-density region, the $\vec{E}_{\parallel}$ contribution falls below zero. This behavior is caused by the evolution of the wakefield. While the electrons are in the accelerating phase, they beam load the wake [34] and reduce the accelerating field. By the time the electrons enter the decelerating portion of the wake, the evolution of the wake due to the photon deceleration ($a_0 \sim 1/\omega$) leads to a higher decelerating gradient. The length over which deceleration occurs is also longer than the acceleration distance. Certainly, as can be seen in figure 7, some of the energy that the electrons had gained from DLA is returned to the wake and possibly to the laser. Thus, figure 7 shows that even beyond the dephasing length, LWFAs where DLA is present can have energies that exceed the dephasing-length-limited energy gain due to the DLA contribution.

## 4. Conclusion

We have used 2D OSIRIS PIC simulations to investigate LWFA with ionization injection of electrons in a quasi-blowout regime where the laser pulse overlaps the trapped electrons. The electron beams produced have maximum electron energies that exceed the dephasing-length-limited energy estimates given by the theory for the ideal blowout regime. The simulations show that DLA can be an additional acceleration mechanism when the LWFA is operating in this regime. The relative contributions to the maximum electron energy from the longitudinal and transverse fields were determined, and it was demonstrated that for a given laser pulse length, the DLA contribution increases with the plasma density as expected. Further, it was found that at high densities, the DLA contribution can be greater than the longitudinal field contribution in determining the maximum energy gain of the electron bunch.

**Acknowledgements**

Experimental work supported by DOE grants DE-FG02-92-ER40727 and DE-SC0010064 and NSF grant PHY-0936266. Simulation work done on the Hoffman2 Cluster at UCLA and on NERSC. This publication was created with government support under and awarded by DoD, Air Force Office of Scientific Research, National Defense Science and Engineering Graduate (NDSEG) Fellowship 32 CFR 168a (Shaw), NSF Graduate Fellowship DGE-0707424 (Shaw), and NSERC Canada PGS-D (Vafaei-Najafabadi).